%% file: main.tex
\documentclass[10pt,conference]{IEEEtran} 
\IEEEoverridecommandlockouts
\usepackage{cite}
\usepackage{amsmath,amssymb,amsfonts}
\usepackage{algorithmic}
\usepackage{graphicx}
\usepackage{textcomp}
\usepackage[x11names]{xcolor}
\usepackage{soul}
\usepackage{enumitem}

\usepackage[hidelinks]{hyperref}

\usepackage{listings}
\usepackage[cachedir=minted-cache]{minted}
\setminted{fontsize=\scriptsize,numbersep=6pt}

\usepackage{array}
\usepackage{booktabs}
\usepackage{multirow}
\usepackage{float}

\usepackage[utf8]{inputenc}
\usepackage[font=footnotesize]{caption}
\usepackage{subcaption}
\usepackage{xspace}

\usepackage{transparent}

\usepackage{titlesec}
\titlespacing*{\subsubsection}{0pt}{0.1\baselineskip}{0.2\baselineskip}
\titlespacing*{\subsection}{0pt}{0.2\baselineskip}{0.2\baselineskip}
\titlespacing*{\section}{0pt}{0.4\baselineskip}{0.2\baselineskip}

\input{cmds}

\def\BibTeX{{\rm B\kern-.05em{\sc i\kern-.025em b}\kern-.08em
    T\kern-.1667em\lower.7ex\hbox{E}\kern-.125emX}}

\begin{document}

\title{\toolName: An Instrumentation-based Dynamic Analysis Framework for Rust}

\input{parts_authors}

\maketitle

\input{parts_abstract}


\input{parts_s1-introduction}

\input{parts_s2-overview}

\input{parts_s3-interface}

\input{parts_s4-instrumentation}

\input{parts_s5__-evaluation}

\input{parts_s7-related}

\input{parts_s9-conclusion}

\input{parts_s_-references}

\end{document}

%% file: cmds.tex
\newcommand{\toolName}{\textsc{Leaf}\xspace}

\newcommand{\code}[1]{\texttt{\detokenize{#1}}}

\newcommand{\mysubsubsection}[1]{\noindent\textbf{#1}}

%% file: parts_authors.tex
\author{
\IEEEauthorblockN{Mohammad Omidvar Tehrani\IEEEauthorrefmark{1}, Marco Gaboardi\IEEEauthorrefmark{2}, Nick Sumner\IEEEauthorrefmark{1}, Steven Y. Ko\IEEEauthorrefmark{1}} \\
\IEEEauthorblockA{
\IEEEauthorrefmark{1}\textit{School of Computing Science}, \textit{Simon Fraser University},Vancouver, Canada \\
}
\IEEEauthorblockA{
\IEEEauthorrefmark{2}\textit{Department of Computer Science}, \textit{Boston University},
Boston, USA \\
}
}

%% file: parts_abstract.tex
\begin{abstract}
  This paper presents \toolName, an instrumentation-based dynamic analysis framework for Rust.
  Although Rust has grown rapidly in recent years, the landscape of program analysis tools for Rust
  is still in relatively early stages. One notable gap is the lack of a general-purpose dynamic
  analysis framework that can support different analysis tasks. \toolName aims to fill this gap by
  providing a Rust-native framework for analyzing Rust programs at runtime. Rust provides rich
  semantic information through its ownership model, type system, memory model, and compiler-level
  representation. Therefore, \toolName focuses on how to make this information available to dynamic
  analyses. In particular, \toolName captures \emph{MIR-level semantic information}, augments it
  with \emph{runtime facts}, and delivers it to analyses as an \emph{event stream} through
  \emph{Dynamic MIR (DMIR)}, an event-driven programming interface. Through three substantial
  dynamic analyses---a concolic executor, a Rust-specific sanitizer, and a control-flow tracer---we
  demonstrate the practicality and expressiveness of \toolName. Our evaluation further shows that
  \toolName's compile-time and runtime overhead is meaningful but manageable.
\end{abstract}

%% file: parts_s1-introduction.tex
\section{Introduction}

Despite Rust's rapid growth as a safe programming language, its program analysis ecosystem is still
relatively young. Rust has strong support for linting~\cite{clippy}, fuzzing~\cite{cargo_fuzz}, bounded model
checking~\cite{kani}, and deductive verification~\cite{noauthor_creusot_nodate}, but one notable missing piece is a
\emph{general-purpose framework for dynamic analyses}. Such a framework would enable a variety of
analyses, including taint analysis, tracing, coverage analysis, and symbolic execution. More
established languages already have well-designed frameworks that provide reusable infrastructure for
building dynamic analyses, such as DynaPyt~\cite{eghbali_dynapyt_2022} for Python,
Wasabi~\cite{lehmann_wasabi_2019} for WebAssembly, DiSL~\cite{marek_disl_2012} for Java, and
Dyninst~\cite{williams_dyninst_2016}, Pin~\cite{luk_pin_2005}, and
DynamoRIO~\cite{bruening_infrastructure_2003} for native binaries. These frameworks handle the
low-level mechanics of code instrumentation and data collection, allowing analysis developers to
focus on the logic of their analyses.

Since these frameworks are designed for other languages or lower-level binaries, analysis developers
cannot directly use them to build dynamic analysis tools for Rust. More importantly, their
abstractions do not expose the language semantics that Rust-specific dynamic analyses need. Rust's
ownership model, for example, enforces strict rules about how memory is accessed, shared, moved, and
dropped. These semantics are central to Rust programs, but they are not naturally represented by
frameworks designed for languages without ownership or by frameworks that operate only at the binary
level. This points to the need for a Rust-native dynamic analysis framework that can support a wide
range of analyses while preserving Rust's language-level semantics.

This paper presents \toolName, an instrumentation-based dynamic analysis framework for Rust. The
central design challenge of \toolName is \emph{how to make Rust's rich semantic information
available} so that dynamic analyses can use it. Rust's ownership model, type system, memory model,
and compiler-level representation expose information that can be valuable for dynamic analyses, but
this information is not automatically available at runtime in a form that analyses can use. Building
a Rust-native framework therefore requires answering three questions. First, what Rust semantic
information should the framework capture? Second, how should this information be delivered to
analyses during program execution? Third, how should the framework present this information through
an abstraction that supports different types of analyses without exposing unnecessary low-level
details?

\toolName answers these questions by centering its design on one goal---delivering \emph{MIR-level
semantic information}, augmented with \emph{runtime facts}, as an \emph{event stream} to dynamic
analyses. Concretely, this means the following design choices. First, \toolName uses MIR (Mid-level
Intermediate Representation) to capture Rust's rich semantics, such as ownership, lifetimes, types,
and control flow. This is a natural choice. MIR is designed to capture these semantic details so
that the Rust compiler can perform its analysis, e.g., borrow checking. Second, since MIR is a
static representation used at compile time, it does not capture runtime facts that are available
only during execution. Examples include concrete types for generics, variable values and memory
addresses, and the branches taken. \toolName uses instrumentation that statically inserts operations
that capture MIR and runtime information. For this, we develop techniques to effectively perform
this instrumentation in \toolName. Third, since different analyses require different subsets of this
information, \toolName also supports \emph{configurable} instrumentation, allowing analyses to
control which functions and kinds of information are collected. Fourth, \toolName develops an
event-driven programming model called \emph{DMIR (Dynamic MIR)} that presents the MIR and runtime
information to dynamic analyses. DMIR faithfully represents MIR and runtime information so that
analysis developers can focus on their analysis logic rather than handling the low-level mechanics
that produce the event stream.

To demonstrate the power of \toolName, we implement three dynamic analyses with significant
complexity---a concolic executor, a Rust-specific sanitizer, and a control-flow tracer. Our concolic
executor involves 11,749 lines of code and performs full concolic execution with the Z3 SMT
backend~\cite{de_moura_z3_2008}. Our Rust-specific sanitizer detects use-after-drop and double-drop
errors in Rust programs with 1,746 lines of code. Our control-flow tracer uses a built-in feature of
\toolName, enabling full control flow tracing with only 646 lines of code. Together, these analyses
demonstrate \toolName's expressiveness and practicality.

In addition, we quantify the compile-time and runtime overhead of \toolName using eight popular Rust
crates from \texttt{crates.io}, as well as the Rust compiler performance benchmark suite. Our
results show that the overhead is meaningful but manageable, and consistent with the costs expected
from dynamic analysis frameworks.

In summary, we make the following contributions:
\begin{itemize}[leftmargin=*, itemsep=0pt, topsep=0pt]
    \item We develop DMIR, an event-driven interface that delivers MIR and runtime
      information to dynamic analyses.
    \item We develop instrumentation techniques for capturing and delivering Rust semantic information through DMIR.
    \item We implement three substantial dynamic analyses, a concolic executor, a Rust-specific
      sanitizer, and a control-flow tracer.
    \item We evaluate \toolName on popular crates and Rust compiler benchmarks, quantifying its
      compile-time and runtime costs.
\end{itemize}

The \toolName implementation consists of 26,739 lines of code. The code for \toolName and the three
use cases, as well as instructions on how to run them, are available via our artifact submission at
\url{https://github.com/sfu-rsl/leaf}.

%% file: parts_s2-overview.tex
\section{Overview}

\begin{figure}
    \centering
    \includegraphics[width=\columnwidth]{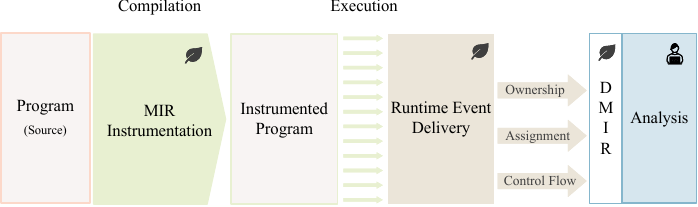}
    \caption{Overview of \toolName. Analysis developers implement callbacks over DMIR events, while
    \toolName instruments MIR, collects MIR information and runtime facts, and delivers them to the
    callbacks during execution.}
    \label{fig:overview}
    \vspace{-1.5em}
\end{figure}

The central objective behind \toolName's design is to faithfully deliver MIR's rich semantics,
augmented with runtime facts, as an event stream for dynamic analyses. From an analysis developer's
point of view, using \toolName means implementing callbacks in \emph{Dynamic MIR (DMIR)}, the
event-driven programming interface \toolName provides. More important, \toolName captures the rich
semantics provided by MIR in these callbacks.

For example, \code{let y = &mut x;} shows an example of \emph{mutable borrow} in Rust, which creates
a mutable reference to \code{x} and stores it in \code{y}. At runtime, this triggers an assignment
event in \toolName. A developer implements the corresponding assignment callback to handle this
event in their analysis. The callback receives the MIR-level semantics of the assignment, including
that the right-hand side is a mutable borrow of \code{x}. In other words, the analysis observes the
operation as a Rust-level assignment with borrowing semantics, rather than as a simple write
operation. We discuss the details in \autoref{sec:dmir}.

To realize this interface, \toolName collects MIR information and runtime facts, and delivers them
in the form expected by DMIR. \toolName does this in two steps. First, it instruments the program at
the MIR level by inserting operations that collect the information needed for DMIR events. Second,
at runtime, \toolName organizes the collected information into DMIR's structure and delivers it to
the corresponding analysis callbacks. \autoref{fig:overview} shows the overall flow of \toolName. We
describe this realization in \autoref{sec:realizing-dmir}.

\toolName also provides utilities to support analysis tasks. These include
a call flow manager to manage data and control transfer between function calls,
a shadow memory for tracking memory state, and a non-blocking logger for recording runtime information.
Analysis developers can use them as needed to assist with their core analysis logic.

The following sections describe these components and evaluate how well they support
practical Rust dynamic analyses.

%% file: parts_s3-interface.tex
\section{Dynamic MIR (DMIR)}
\label{sec:dmir}

\emph{Dynamic MIR (DMIR)} is the event-driven programming interface \toolName provides to analysis
developers. The goal of DMIR is to faithfully present MIR's rich semantics, augmented with runtime
facts, as an event stream for dynamic analyses. To do so, DMIR preserves the MIR constructs, such as
types, statements, ownership, lifetime, function calls, control flow, etc. However, DMIR exposes
these constructs in a different form than MIR---while MIR is a static, compile-time representation of a
program, DMIR presents the same constructs as they occur during program \emph{execution}. This
distinction between static representation and runtime event stream is central to DMIR's design. We therefore first provide a brief overview of MIR
and then describe DMIR further.

\begin{table*}[ht!]
    \centering
    \caption{MIR constructs and related callbacks in DMIR corresponding to Rust execution events.}
    \input{res_tbls_dmir_sigs}
    \label{tab:events-and-callbacks}
    \vspace{-1.5em}
\end{table*}

\mysubsubsection{Background on MIR:}
MIR is a compiler-internal, static representation of a Rust program, including
variables, operations, basic blocks, types, and other program constructs.
Although \texttt{rustc} can emit a human-readable form of MIR, this form is only
informational. We use it in examples for readability, but MIR itself is
represented internally as compiler data structures, just as DMIR is exposed
through data structures and callbacks (as described later).

From the perspective of program analysis, MIR is often \emph{the} source of the rich semantics of
Rust, such as operations, operands, ownership, lifetime, control flow, etc. MIR is designed to
provide the semantics of Rust so that the Rust compiler can perform its analyses such as borrow
checking~\cite{mir_rustcdev}. This information is difficult to recover from lower-level
representations \cite{jung_miri_2026, min_erasan_2024}, so most program analysis tools for Rust
perform MIR-level analysis. DMIR preserves MIR information for the same reason.

\mysubsubsection{DMIR Overview:} DMIR exposes MIR-level information during execution, augments it
with runtime facts, and delivers the result as an event stream. Each event in this stream gives the
analysis three kinds of information---the event being executed, the static MIR information
associated with it, and the runtime facts observed during execution.

First, DMIR delivers events triggered at MIR-level execution points. Examples include an assignment,
a function call, and a conditional jump. An analysis developer implements a dynamic analysis by
handling these events via \toolName-defined callbacks. \autoref{tab:events-and-callbacks} shows event examples and the corresponding callbacks, which we explain further in
\autoref{sec:events}.

Second, each event carries static MIR information about the construct being executed. For example,
an assignment event identifies the destination and the operand being assigned, and a function call event
identifies the call site and callee information. This information comes from MIR and allows analyses to interpret
runtime behavior with MIR-level semantics.

Third, DMIR augments this MIR information with runtime facts collected during execution. Examples
include memory addresses, the branch actually taken, the concrete target of a dynamic call, and the
concrete type of a generic. These facts are unavailable in static MIR alone, but are necessary for
dynamic analyses that need to reason about concrete execution.

Together, these three kinds of information form the DMIR event stream---a runtime stream of events
that preserves MIR-level semantics while adding the execution facts needed by dynamic analyses. The
rest of this section presents DMIR in more detail, starting with events.

\subsection{Events}
\label{sec:events}

DMIR represents program execution as a stream of events. Each event corresponds to the execution of
a MIR construct, such as an assignment, a function call, a branch, a drop, or a lifetime operation.
An analysis developer implements a dynamic analysis by handling these events through
\toolName-defined callbacks. In this sense, DMIR is not a low-level trace of instructions or memory
accesses. Rather, it is a MIR-level trace of the program's execution.

\autoref{tab:events-and-callbacks} shows representative DMIR events and the corresponding callbacks
that analysis developers implement to handle them. To avoid requiring detailed knowledge of Rust
syntax, we simplify the callback signatures in the table. The examples include language-agnostic
events, such as assignments and function calls, as well as Rust-specific events, such as ownership
events (move, borrow, and drop) and lifetime events (live and dead).
DMIR, however, is not limited to these examples---it exposes callbacks for all MIR-level execution
events. In total, DMIR defines 50 callbacks.

\autoref{fig:callback-example} illustrates one such callback. The example shows an assignment
callback that prints information about a binary operation. The callback receives the operation, the
operands, and the destination as parameters. These parameters are important because they preserve
the MIR-level description of the assignment rather than reducing it to a lower-level read, write, or
instruction event, thereby losing the semantics. We discuss how \toolName preserves MIR-level
semantics further next.

\subsection{MIR Semantics and Runtime Facts}

Each DMIR event carries two kinds of information. The first is MIR information, which describes the
construct being executed. The second is runtime facts, which are observed while executing that
construct. This combination is central to DMIR's design as MIR information preserves what the
operation is, while runtime facts capture what happened during this particular execution.

For example, consider \autoref{fig:callback-example} once again. The callback receives two operands
and the destination. These are not simple variable names or raw values. They are DMIR operands that
preserve the MIR-level structure of the operation while also exposing runtime facts. As explained
later in this section,
MIR represents memory locations using {\emph{places}} and describes how they are accessed as {\emph{operands}}.
For example, an operand can describe whether the operation moves a field from a struct, copies an item from a slice, or copies a value pointed to by
a reference. DMIR preserves this place-level information and makes it available to the analysis.

The example illustrates this dual role. The first print statement uses the operands and the destination place
to describe the operation being executed (lines 7). This reflects the MIR-level view of the assignment, i.e.,
what operation is occurring and which places are involved (line 15). The second print statement queries
runtime facts from the same objects, such as the destination address and the values observed for the
two operands (lines 10 and 16). Thus, a DMIR callback gives the analysis both the MIR structure of the event and the
concrete facts observed during execution.

\begin{figure}[t]
\begin{minted}{rs}
impl AssignmentHandler for ExampleAssignmentHandler {
  fn binary_op(self,
    id: AssignmentId, dest: Self::Place, op: BinaryOp,
    first: Self::Operand, second: Self::Operand,
  ) {
    println!("{dest} = {first} {op} {second}");
    println!("{} <- {}, {}, {}",
      dest.addr(), op.name(), first.val(), second.val(),
    );
  }
}
// Output:
// Local(6) = Moved(Local(5).Field(1)) + Const(5, u32)
// 0x7777ff0 <- Add, 10u32, 5u32;
\end{minted}
    \caption{Example of implementing callback for assignments.}
    \label{fig:callback-example}
    \vspace{-1.5em}
\end{figure}

The rest of this section illustrates this design through important cases, including
places, control flow, intrinsics, static identifiers, and incomplete event streams.

\mysubsubsection{Places:} Places are MIR's representation of program memory. A place represents a
memory location starting from a local variable and applying a chain of \emph{projections}. For example,
dereferencing a reference is represented as the local variable for the reference plus a
dereference projection. Accessing a struct field is represented as the place for the struct plus a
field projection.

A DMIR place preserves this MIR structure and adds the runtime facts needed to interpret it during
execution. In particular, DMIR reports the concrete type and address of each component in the place.
Concrete types matter because MIR may contain generic parameters such as \code{T}, whereas execution
uses the concretized type such as \code{u8}. Addresses matter because unsafe Rust can reach the same
memory through raw pointers that are visible only at runtime. We return to the delivery of this
information in \autoref{sec:realizing-dmir}.

An important design point in DMIR is that places and operands are not tied to a fixed
representation. Instead, DMIR allows analyses to choose their own representations early, so later
callbacks can already use the form that best fits the analysis. When a place is constructed, DMIR
passes the MIR place together with its runtime facts to the analysis, and the analysis returns the
representation it wants to use. The same is true for operands created from moved or copied values.
After this point, DMIR treats these analysis-defined representations as black boxes---it simply
passes them to later callbacks. The types in \autoref{fig:callback-example} (lines 3 and 4)
illustrate this design. \code{Self::Place} and \code{Self::Operand} are analysis-defined, custom
representations of places and operands. \toolName passes these to the callback rather than
predefined forms.

\mysubsubsection{Control Flow:} MIR represents two kinds of control flow. One is regular branching,
as in conditionals and loops. The other is function calls. A function call in MIR does not always
have a single target, as it may depend on (i) generic type parameters, (ii) function pointers, or
(iii) dynamic dispatch, such as calling a method of a trait object~\cite{trait_object_rustref}. In
all cases, the target is resolved by the time the program executes.

Thus, DMIR reports the actual path taken during execution. For conditional branches, DMIR reports
each branch target taken. For function calls, DMIR reports the entrance into and return from each
function. In this way, the analysis sees the control flow structure from MIR together with the path
taken in this execution.

\mysubsubsection{Intrinsics:} Rust provides \emph{intrinsics} for operations that require special
compiler support, e.g., raw-memory operations and atomics. In MIR, these intrinsics appear similar
to function calls, but they do not behave like regular calls. Instead, the compiler gives them
special meaning, such as performing a low-level operation directly.

If DMIR exposed intrinsics only as function calls, analysis developers would have to recover this
special meaning themselves. Therefore, DMIR treats intrinsics as first-class operations and reports
the operation they perform, together with the relevant runtime facts. This lets analyses handle
intrinsics according to their MIR-level meaning, rather than as regular function calls.
\autoref{fig:intrinsics-example} shows one such operation, {\code{rotate_left}}, which is modeled
uniformly as regular binary operations.

\begin{figure}[t]
\begin{minted}{text}
// MIR
_4 = intrinsics::rotate_left::<u8>(copy _1, const 2_u32);
// DMIR
assign_to(place({local: 4}))
  .binary_op(copy_of(place({local: 1})), cons(2u32), RotateL);
\end{minted}
    \caption{DMIR treats MIR instrinsics as first-class operations.}
    \label{fig:intrinsics-example}
    \vspace{-1.5em}
\end{figure}

\mysubsubsection{Static Identifiers:}
DMIR includes static identifiers in runtime events so that analyses can map each event back to the
MIR construct that produced it. These identifiers name static MIR locations, not individual dynamic
executions. In other words, if an assignment runs multiple times in a loop, each runtime event
carries the same static identifier for the assignment.

When available, DMIR reuses MIR's structure for passing identifiers to callbacks. For example,
function-entry events carry MIR's function identifier, and control-flow events carry the index of
the basic block that they terminate. DMIR additionally defines unique identifiers for assignments
within their function. Together, these identify the events, and allow analyses to connect the
runtime stream with compile-time information such as source locations, MIR metadata, or
static-analysis results.

\mysubsubsection{Incomplete Event Streams:} DMIR does not always contain every event or every piece
of information from the actual execution. This can happen when some code cannot be instrumented,
such as external functions, or a developer's configuration (described in
\autoref{sec:configurations}) instructs \toolName to omit certain code. In these cases, DMIR still
tries to preserve the shape of the MIR-level event stream, while making the missing information
explicit to the analysis.

DMIR does this in two ways. First, when part of an event is omitted, DMIR uses an opaque value,
\emph{Some}, in place of the missing information. For example, if an assignment is included but its
operands are not, the event is still delivered, but the missing operands are represented as
\emph{Some}. This allows the analysis to know that the assignment happened, even if they do not need
all the details.

Second, DMIR uses function IDs to make gaps in the event stream visible. When a call occurs, DMIR
reports the function being called. If that function is instrumented, the analysis will also see the
corresponding function-entry event with the same function ID. If the entry event does not appear,
the analysis can tell that execution entered code outside the visible DMIR stream, such as an
external or excluded function.

%% file: res_tbls_dmir_sigs.tex
\setminted{escapeinside=||, fontsize=\scriptsize}
\usemintedstyle{bw}

\def\eventwidth{0.37\textwidth}
\def\mirwidth{0.17\textwidth}
\def\dmirwidth{0.39\textwidth}

\renewcommand{\arraystretch}{1}

\begin{tabular}{ccc}
    \toprule
    Event & MIR & DMIR\\
    \midrule


    \begin{minipage}{\eventwidth}
    \textbf{Assignment:} Assign the result of the right-hand-side operation to a destination $p_{dest}$.
    \end{minipage}
    &
    \begin{minipage}{\mirwidth}
\begin{minted}{text}
|$p_{dest}$| = |$op$|(|$o_{fst}$|, |$o_{sec}$|);
\end{minted}
    \end{minipage}
    &
    \begin{minipage}{\dmirwidth}
\begin{minted}{rust}
AssignmentHandler::binary_op(
  id: AssignmentId, dest: Place,
  op: BinaryOp, first: Operand, second: Operand);
\end{minted}
    \end{minipage}
    \\
    \hline
    
    \begin{minipage}{\eventwidth}
    \textbf{Move:} Move the value at $p$ into an operand.
    \end{minipage}
    &
    \begin{minipage}{\mirwidth}
\begin{minted}{text}
move |$p$|
\end{minted}
    \end{minipage}
    &
    \begin{minipage}{\dmirwidth}
\begin{minted}{rust}
OperandHandler::move_of(moved: Place) -> Operand;
\end{minted}
    \end{minipage}
    \\
    \hline

    \begin{minipage}{\eventwidth}
    \textbf{Borrow:} Assign a mutable or immutable reference of $p$ to $p_{dest}$.
    \end{minipage}
    &
    \begin{minipage}{\mirwidth}
\begin{minted}{text}
|$p_{dest}$| = &|$p$|;
\end{minted}
    \end{minipage}
    &
    \begin{minipage}{\dmirwidth}
\begin{minted}{rust}
AssignmentHandler::by_ref(
  id: AssignmentId, dest: Place,
  place: Place, is_mutable: bool);
\end{minted}
    \end{minipage}
    \\
    \hline

    \begin{minipage}{\eventwidth}
    \textbf{Function Call:} Generates two events (before/after). Showing only the after-callback for brevity.
    \end{minipage}
    &
    \begin{minipage}{\mirwidth}
\begin{minted}{text}
|$p_{res}$| = |$f$|(|$op_{a_0}$|, |$op_{a_1}$|,|$...$|);
\end{minted}
    \end{minipage}
    &
    \begin{minipage}{\dmirwidth}
\begin{minted}{rust}
CallHandler::before_call(
   callee_id: FuncId, loc: BasicBlock,
   callee: Operand, args: [Operand]);
\end{minted}
    \end{minipage}
    \\
    \hline


    \begin{minipage}{\eventwidth}
    \textbf{Drop:} Drop a memory location. Generates two events (before/after).
    \end{minipage}
    &
    \begin{minipage}{\mirwidth}
\begin{minted}{text}
drop(|$p$|);
\end{minted}
    \end{minipage}
    &
    \begin{minipage}{\dmirwidth}
\begin{minted}{rust}
DropHandler::before_drop(place: Place);
DropHandler::after_drop();
\end{minted}
    \end{minipage}
    \\
    \hline

    \begin{minipage}{\eventwidth}
    \textbf{Live/Dead:} Mark a local variable's lifetime as live or dead.
    \end{minipage}
    &
    \begin{minipage}{\mirwidth}
\begin{minted}{text}
StorageLive(|$l$|);
StorageDead(|$l$|);
\end{minted}
    \end{minipage}
    &
    \begin{minipage}{\dmirwidth}
\begin{minted}{rust}
LifetimeHandler::mark_live(place: Place);
LifetimeHandler::mark_dead(place: Place);
\end{minted}
    \end{minipage}
    \\
    \bottomrule
\end{tabular}

%% file: parts_s4-instrumentation.tex
\section{Realizing DMIR}
\label{sec:realizing-dmir}

While the interface of \toolName (DMIR) focuses on \emph{presenting} MIR information and runtime
facts as an event stream, the internals of \toolName focus on the techniques for \emph{delivering}
such information. This requires answering two questions---(i) information \emph{collection}, i.e.,
how to collect the MIR information and runtime facts, (ii) information \emph{delivery}, i.e., how to
deliver this information in the form expected by DMIR. This section presents the design and
implementation techniques used to answer these questions.

\subsection{Information Collection via Probes}
\label{sec:probes}

Our main mechanism for information collection is MIR instrumentation, i.e., we design a compiler
(\texttt{leafc}) that extends the Rust compiler (\texttt{rustc}). \texttt{leafc} inserts
\emph{probes} at all MIR execution points such as assignments, function calls, and conditional
jumps. These probes are essentially function calls that eventually ship the necessary MIR-level and
runtime information to the analysis. Therefore, each probe is supported by a group of surrounding
operations that gather such information. These supporting operations and the probe form MIR blocks
that \texttt{leafc} inserts into the program.

Constructing such MIR blocks faces two challenges. First, it must be simple to generate. Complex
instrumentation logic makes the compiler pass error-prone and difficult to maintain. Second, it must
be conscious of runtime overhead. Faithfully delivering MIR-level information can require
communicating rich structures as these structures can be deeply nested and large. Two good examples
that illustrate this point are places and types. In both cases, a long chain of nesting is possible,
e.g. a list of projections for places, or the layouts of nested fields for types.
Materializing and passing them directly at every event would be expensive. To
address these challenges, \toolName uses five techniques as follows.

The first technique is to use different probes for different variants of an event. When event
variants require different information, \toolName uses separate probes rather than passing the
variant as an additional argument. For example, entering a method of a trait implementation requires
more information than entering a regular function, so \toolName uses separate probes for these two
cases. This keeps each probe signature specific to the information needed for that case. Using a
single probe would need extra arguments that are meaningful only for some variants, which would make
generating the MIR blocks more complicated and possibly cause runtime overhead.

The second technique is the opposite---using one probe for multiple related events. When several
events require the same information, \toolName uses one probe and passes the specific event kind as
an argument. For example, Rust defines separate atomic intrinsics for several binary operations, but
they all need the same type of information. Thus, \toolName merges them into a single probe. This
avoids duplicating probes that would have the same shape and differ only in the operation kind.

The third technique is compositional construction. Some MIR constructs such as places and operands
consist of smaller components that can be passed more conveniently. Thus, \toolName uses a
\emph{helper} call that constructs one component at a time and returns a temporary handle to it.
Later helper calls can use these handles to construct larger components. For example,
\autoref{fig:place_ref_example} shows how the source place \code{(*_12)[_4]} is constructed by first
creating a handle for \code{_12} (line 5), then a handle for dereferencing it (line 6), and finally
a handle for indexing it with \code{_4} (lines 7 and 8). The copied operand is then constructed from
this place (line 10). The event probe receives only the final handles needed for the assignment
event.

The fourth technique is to use static references. Some information is too large
to pass directly through probes. \toolName therefore dumps such information into compile-time
databases and passes only numeric identifiers through probes. A prominent instance is type
information, which includes details such as memory layout, fields for structures, or pointees for
reference and pointer types. Instead of passing all this information as arguments, \toolName stores
this in the type database that assigns an ID to a type, and pass the type IDs in the probes. A
unique ID is also passed for assignments, which enables looking up information like their accurate
location in the source MIR during runtime.

The fifth technique is inlining and encoding. When a particular kind of temporary handle (created by
compositional construction) is used frequently, \toolName encodes the relevant information directly
into the handle. This avoids additional helper calls or database lookups. For example, a place that
refers to a local without projections does not need to be constructed through a separate helper
call. Therefore, \toolName encodes the local directly into the temporary handle.
This helps with the overhead of compilation as fewer
probes are inserted, and with the performance at runtime as the handle or the id does not need to be
looked up.

\begin{figure}[t]
\begin{minted}{text}
_10 = copy (*_12)[_4];

_321 = ref_place_local(10_u32);

_328 = ref_place_local(12_u32);
_333 = ref_place_deref(_328);
_338 = ref_place_local(4_u32);
_345 = ref_place_index(_333, _338);

_352 = ref_operand_copy(_345);

assign_use(_321, _352);
\end{minted}
    \caption{Handles used to pass the information of two places and the operand in the assignment event.}
    \label{fig:place_ref_example}
    \vspace{-1.5em}
\end{figure}

\subsection{Information Delivery via Probe-to-DMIR Adapter}
\label{sec:PDA}

Once probes and their MIR blocks collect the MIR and runtime information, it
needs to be organized into the form expected by DMIR. This is the job of the
Probe-to-DMIR Adapter (PDA). The need for this adapter comes from the mismatch between probes and
DMIR. Probes are designed for instrumentation and therefore use a lower-level format. Specifically,
there are three cases among the five cases mentioned in \autoref{sec:probes} that require special
handling---a probe may (i) represent several related events using an additional argument, (ii)
encode information into compact values, and (iii) receive information via temporary handles produced
by compositional construction. DMIR, on the other hand, presents this information as a structured
event stream to the analysis.

PDA bridges this gap through three tasks. First, when several related events are multiplexed through
one probe, PDA uses the event argument to identify the event and reconstruct the corresponding DMIR
event. Second, when probe arguments encode information, PDA decodes them back into the structures
expected by DMIR. Third, PDA manages compositional constructions internally and resolves those
handles before invoking the DMIR callback. In \autoref{fig:place_ref_example}, the helper probes
construct the destination place, the source place, and the copied operand before the assignment
probe is called. The assignment probe receives only two handles, one for the destination and one for
the operand. PDA implements the helper probes, collects the information passed in them, and presents
the information to the DMIR assignment event.

These tasks allow probes to remain simple and efficient while allowing DMIR to present
MIR-level information in the form expected by DMIR.
However, not all information needed by DMIR can be delivered by this general probe-to-DMIR
adaptation alone. Some information requires specialized handling because the form available during
MIR instrumentation differs from the form needed during execution. We discuss two such cases
next---concrete types and function identifiers.

\subsection{Types}
\label{sec:types-realization}

DMIR requires concrete types at runtime, but the MIR can still contain generic types. PDA addresses
this gap with a generic helper probe that carries the same type parameters as the MIR body where it
is inserted. As a result, when the compiler later generates concrete code for that body (via
monomorphization), it also generates a concrete version of the helper probe. This allows PDA to
observe the concrete type used in that execution and deliver it to DMIR.

\subsection{Function Identifiers}

MIR assigns a static ID for each function. However, a generic function, a function-pointer call, or
a trait-object method call may point to different concrete functions at runtime.
DMIR therefore augments MIR's static identifier with a dynamic one to identify functions precisely.

The dynamic identifier used for DMIR is raw address of the function that is actually called and
entered. The address is directly available for function pointer and monomorphized generic functions.
For methods of trait objects, a pair of the address of object and the index of method is used.

\subsection{Configurations}
\label{sec:configurations}

The mechanisms above describe how \toolName delivers MIR-level information to DMIR. In practice, an
analysis may not need all of this information. In addition, delivering every event and every
component can be expensive. Thus, \toolName allows analyses to configure what information to
collect.

This is done by using rules over MIR constructs. Each rule tells \toolName whether to include or
exclude a construct. A rule has two parts. The first part identifies the MIR bodies the rule applies
to. The second part identifies the MIR constructs to include or exclude at a fine granularity. \autoref{lst:config_example} shows an example configuration.

\begin{figure}[t]
\begin{minted}{toml}
[[instr_rules.include]]
entity = "place_info"
kind = "address"
loc = { def_path = "my_imp::raw::*" }
[[instr_rules.exclude]]
entity = "body"
all = [ { crate = { is_external = true } } ]
\end{minted}
\caption{Sample configuration to exclude all external bodies (in dependencies) from instrumentation, and augment places with their raw addresses in module \code{my_imp::raw}.}
\label{lst:config_example}
\vspace{-1.5em}
\end{figure}

\subsection{Dependency Instrumentation}

A Rust program usually depends on other crates (external libraries). To produce a complete event
stream, \toolName must instrument those crates as well. One possible approach is to instrument each
dependency when it is first compiled. Since dependencies are compiled before the target crate, the
target program would then receive the already-instrumented versions of those dependencies when it is
built.

However, this early-instrumentation approach has several drawbacks. Because the target program is
not yet known, \toolName may instrument dependency code that the target program never uses. It may
also insert probes into functions that the compiler would otherwise inline into the target program.
Finally, it must choose an instrumentation configuration before knowing what the analysis needs for
this particular program.

\toolName therefore takes a different approach. During the build of the target crate, it forces the
compiler to recompile the MIR bodies of dependencies that participate in the build, and runs its MIR
instrumentation pass during this recompilation. This lets \toolName instrument the target program
and its dependencies in one coordinated build, with the instrumentation configuration chosen for the
current analysis. The cost is additional build effort---large dependencies, including the standard
library, may be rebuilt. As a result, instrumented builds can be slower than normal builds even before accounting
for the inserted probes. We report compilation performance in \autoref{sec:eval_instr}.

%% file: parts_s5__-evaluation.tex
\section{Evaluation}
\label{sec:evaluation}

Our evaluation aims to answer the following three research questions:

\begin{enumerate}[label=\textbf{RQ\arabic*:}, leftmargin=*, itemsep=0pt, topsep=0pt]
    \item Does \toolName\ enable practical specific dynamic analyses?
    \item What are the compilation/runtime costs of MIR instrumentation, and how do 
      configurations affect them?
    \item Does \toolName's instrumentation preserve the original behavior of instrumented
      programs?
\end{enumerate}

To answer RQ1, we implement three dynamic analyses using \toolName. To answer RQ2 and RQ3, we
instrument real-world Rust crates and the Rust compiler performance benchmark under different
configurations. We measure compilation and runtime overhead, as well as compare configurations to
evaluate the effect of configurability (RQ2), and run each crate's test suite to check behavioral
preservation (RQ3).

\input{parts_s5_1-eval_use_cases}

\input{parts_s5_2-eval_instr}

%% file: parts_s5_1-eval_use_cases.tex
\subsection{RQ1: Use Cases}
\label{sec:use-cases}

To answer RQ1, we implement three dynamic analyses using \toolName. These analyses are not toy
examples---our first dynamic analysis, concolic execution for Rust, consists of 11,749 lines of code
and is able to perform full concolic execution on a Rust program. Our
second analysis, a \code{ManuallyDrop} sanitizer, is a unique analysis with 1,746 lines of code that
detects Rust-specific double-drop and use-after-drop errors involving manually managed ownership.
Our third analysis, a control flow tracer, demonstrates the built-in capability of \toolName as it
is able to track complete control flow within a program with just 646 lines of code. Together, these
analyses demonstrate that \toolName's MIR-preserving event stream with runtime facts can support
practical dynamic analyses while preserving the Rust-specific information those analyses require.
The code for all three analyses, as well as the instructions on how to run them, are available via
our artifact submission at \url{https://github.com/sfu-rsl/leaf}.

\begin{figure}[t]
    \begin{subfigure}{\columnwidth}
\begin{minted}{rust}
fn main() {  // Driver
  let idx = 1.mark_symbolic();
  print_from("Héllo", idx);
}
fn print_from(s: &str, idx: usize) {
  if idx >= s.len() { return; }
  let to_print = &s[idx..];
  println!("{to_print}");
}
// The standard library
fn index(self, slice: &str) -> &Self::Output {
  let (start, end) = (self.start, self.end);
  match self.get(slice) {
    Some(s) => s,
    None => super::slice_error_fail(/*...*/),
  }
}
fn get(self, slice: &str) -> Option<&Self::Output> {
  if slice.is_char_boundary(self.start) {
    Some(unsafe { &*self.get_unchecked(slice) })
  } else {
    None
  }
}
fn is_char_boundary(&self, index: usize) -> bool {
  // ...
  (self.as_bytes()[index] as i8) >= -0x40
}
\end{minted}
        \caption{Example program tested by the concolic execution analysis. The program uses the index operation over a string slice. The driver defines an entry point for the execution and instructs the analysis which variables should be tracked symbolically.}
        \label{fig:symex-code}
    \end{subfigure}

    \begin{subfigure}[b]{\linewidth}
\begin{minted}{text}
Constraints:
1. @(print_from)[bb1],                  !{>=(<idx>, 6u64)}
2. @(get)[bb0],                         !{==(<idx>, 0u64)}
3. @(get)[bb5],                         !{>=(<idx>, 6u64)}
4. @(get)[bb7],                          {< (<idx>, 6u64)}
5. @(get)[bb9],
      {>=(({72,195,169,108,108,111}[<idx>] as i8), -64i8)}
6. @(index)[bb1],                             ImpliedBy(5)
---------------------------------------------------------
New Input To Check:
Var1 = 2
\end{minted}
        \caption{Constraints collected by the analysis during the execution, coming from the branches affected by the symbolic index.
        In this example, the constraints are boolean expressions.
        They represents the condition checked and whether it was held or not during runtime.
        }
        \label{fig:symex-constraints}
    \end{subfigure}

    \caption{Original source code, MIR, and explanation of an example Rust function.}
    \label{fig:symex}
    \vspace{-1.5em}
\end{figure}

\mysubsubsection{Concolic Execution:}
Concolic execution \cite{sen_cute_2005} is a popular technique for finding bugs in a program. It
runs the program with a concrete input while also tracking symbolic expressions, i.e., formulas that
describe how program values depend on that input. When execution reaches a branch that depends on
the input, the analysis records the corresponding branch condition called a
\emph{path constraint}. The set of path constraints collected during one execution forms a
\emph{path condition}. The analysis can then negate one constraint in this path condition (i.e., the
constraint that corresponds to one branch), use an SMT solver (Z3~\cite{de_moura_z3_2008}) to find a
new input that satisfies the modified condition, and run the program again to explore a different
path. This process continues until the analysis finds a target behavior, e.g., a crash, and reports
the input that triggered it.

Implementing concolic execution requires four kinds of support. First, the analysis needs data flow
events to propagate symbolic expressions through assignments, operands, and memory accesses. Second,
it needs control flow events to record branch decisions and construct path constraints. Third, it
needs call flow information to carry symbolic values across function calls and returns. Fourth, for
Rust, it needs MIR-level information such as places, projections, concrete types, memory layout, and
enum discriminants to interpret Rust operations precisely. \toolName provides these through DMIR.

\autoref{fig:symex-code} shows an example. The function \code{print_from} takes a string and an
index (lines 5-9), checks whether the index is within the string length (line 6), and then uses the
index in a slicing operation. A {\emph{driver}} used to set up the concolic execution calls this
function with a fixed string containing a non-ASCII character (é) and marks the index (\code{idx}) as symbolic, i.e., it instructs our
concolic executor to track the data flow from \code{idx}. Running the instrumented program produces
the formula for the path constraints in \autoref{fig:symex-constraints}.
The first constraint corresponds to the guard in {\code{print_from}}, i.e., the index is not greater than the length of the string (1).
Constraints 2-4 belong to the internal checks in {\code{is_char_boundary}} and {\code{get}}, which check if the index is at the boundaries of the string.
Constraint 5 corresponds to the core condition that checks whether the index falls between
character boundaries, which later implies the result of the branch in {\code{index}} (lines 13-16).
Solving these constraints produces a value for \code{idx} that passes the initial bounds check but
falls between UTF-8 character boundaries, causing execution to reach \code{slice_error_fail} in
\code{index} and panic.

This example uses all four forms of support. First, the symbolic index flows through assignments,
operands, arguments, and return values, so the analysis relies on DMIR's data flow and call flow
events. Second, the bug is exposed by changing a branch decision, so the analysis relies on DMIR's
branch events to construct path constraints. Third, the index is used inside a MIR place that
selects from the string buffer, so the analysis relies on DMIR's place structure to distinguish the
slice base from the index projection. Together with concrete type and memory address information,
this allows the analysis to construct an expression over the entire list of bytes that the symbolic
index can potentially select. Fourth, the panic path depends on the enum-based control flow inside
standard library code---the final branch in \code{index} depends on the enum value returned by
\code{get}, which is itself determined by earlier branches inside the function. DMIR's stable event
identifiers allow the analysis to connect these runtime branch decisions with statically extracted
control dependency information \cite{mccamant_quantitative_2008}.

This use case shows that a significant concolic executor can be built and used to
find realistic bugs using \toolName. The key is not only that \toolName reports runtime events, but that these
events preserve MIR-level semantics. Place structures, concrete type information, memory
layout, and control-dependency information allow the analysis to reason precisely about 
execution without reimplementing these semantics itself.

\begin{figure}[t]
\begin{minted}{rust}
#[derive(Debug)]
pub struct MyOption<T> {
  value: ManuallyDrop<T>, // Dropped if is_some is false.
  is_some: bool,
}
impl<T> MyOption<T> {
  pub fn new(value: T) -> Self {
    Self {value: ManuallyDrop::new(value), is_some: true}
  }
  pub fn change_to_none(&mut self) {
    if self.is_some {
      self.is_some = false;
      unsafe { ManuallyDrop::drop(&mut self.value); }
    }
  }
}
fn main() {
  let mut option = MyOption::new(String::from(""));
  option.change_to_none();
  println!("{:?}", option);
}
\end{minted}
    \caption{Example of use-after-drop.}
    \label{lst:mdsan}
    \vspace{-1.5em}
\end{figure}

\mysubsubsection{\code{ManuallyDrop} Sanitizer:}
\code{ManuallyDrop} is a Rust wrapper that allows programmers to take manual control over dropping a
value. Normally, Rust automatically drops each value exactly once when the value goes out of scope.
With \code{ManuallyDrop}, the programmer may call its unsafe \code{drop} method directly, but must
ensure that the value is dropped at most once and is not accessed after it has been dropped.
A violation can cause undefined behavior~\cite{md_rustdocs}.

\autoref{lst:mdsan} shows an example with a use-after-drop problem that our sanitizer detects.
\code{struct MyOption} (lines 1-6) uses \code{is_some} to remember whether the value inside \code{ManuallyDrop}
is still valid. When \code{is_some} becomes false, the wrapped value has already
been dropped and should no longer be used. The method \code{change_to_none} (lines 11-16) follows
this rule---it updates \code{is_some} and then drops the wrapped value (lines
13-14).

The problem appears when the value is printed. The type derives Rust's built-in \code{Debug} trait
(line 1), which asks the compiler to generate printing code automatically. That generated
code prints each field of the type, including the \code{ManuallyDrop} field. As a result, printing
the value (line 22) accesses the wrapped value after \code{change_to_none} has already
dropped it. Our sanitizer detects this use-after-drop.

The sanitizer is a MIR-level data flow analysis for ownership state. It tracks each relevant
\code{ManuallyDrop} value in shadow memory with a flag for whether the wrapped value has
already been dropped. The state begins when a \code{ManuallyDrop} value is created or assigned,
propagates through copies and moves, and ends when the corresponding storage becomes dead. The
sanitizer reports an error if a later borrow or copy accesses a wrapped value whose dropped flag is
set.

\toolName provides the needed information through DMIR's MIR-level ownership events. It delivers
copies, moves, drops, borrows, and lifetime events together with the MIR places and concrete types
involved. This allows the sanitizer to identify \code{ManuallyDrop} values and the wrapped values
inside them, update their shadow state on ownership events, and remove state when the lifetime event
\code{StorageDead} indicates that a place is no longer live. Because these events preserve MIR
ownership structure, the sanitizer can express the check directly over Rust's ownership operations.

\mysubsubsection{Control Flow Tracer:}
A control flow trace records the functions called and the branches taken during
execution. Such a trace is useful for debugging, profiling, and coverage
measurement. We implement a control flow tracer to show that such an
analysis can be built directly from \toolName's event stream.

The tracer relies on DMIR events for function calls, entries, returns, exits, and branches. These
events carry MIR-level identifiers, so the trace can report call sites, return points, basic blocks,
and branch decisions in terms of MIR rather than low-level instructions. The tracer forwards these
events to \toolName's runtime recording utility, which emits the execution trace. This required only
646 lines of analysis code.

This use case also demonstrates \toolName's configurability. Since the tracer does not need data
flow information, its configuration excludes data-related constructs such as places, operands, and
assignments. Location filters can further restrict tracing to selected modules.
\autoref{sec:eval_instr} reports how these configurations affect instrumentation cost.

%% file: parts_s5_2-eval_instr.tex
\subsection{RQ2: Compilation/Runtime Costs and Configurability}
\label{sec:eval_instr}

\begin{table}[t]
    \caption{Instrumentation configurations used in evaluation.}
    \centering 
    \input{res_tbls_tab_eval_configs}
    \label{tab:eval_configs}
    \vspace{-1.5em}
\end{table}

\begin{table*}[!ht]
    \centering
    \setlength{\tabcolsep}{2pt}
    \centering
    \caption{Compilation costs of instrumentation using the configurations measured in terms of the code generation time and size of the output binary program.}
    \scriptsize
    \label{tab:eval_compilation}
    \input{res_tbls_tab_eval_compilation}
    \vspace{-1.5em}
\end{table*}

\toolName exposes MIR-level semantics and runtime facts as an event stream for dynamic analyses.
This design is inherently costly~\cite{lehmann_wasabi_2019, poeplau_symbolic_2020}---at compile
time, \toolName must perform the tasks described in \autoref{sec:realizing-dmir}, including
inspecting MIR, constructing and inserting instrumentation blocks, and generating probes. At
runtime, the instrumented program must execute those probes and deliver structured information such
as places, projections, concrete types, function identifiers, memory addresses, and control flow
decisions.

In general, dynamic tools for Rust, such as Miri, also incur substantial overhead (as we show later
in this section). This is because they operate over MIR-level program information. RQ2 therefore
quantifies the practical cost of dynamically exposing Rust's rich MIR-level information through
\toolName, in terms of both compilation overhead and runtime overhead.

Our measurements use a machine equipped with an Intel i7-12700K (3.60-4.90 GHz)
and 96GB of RAM.

\mysubsubsection{Compilation Cost:}
We measure the compilation cost of \toolName across three dimensions---target crate, build profile,
and instrumentation configuration. We use eight widely used Rust crates drawn from popular
\texttt{crates.io} categories. The selected crates cover different sizes and functionalities and
include test suites, which we use for RQ3. They include, for example, \emph{Wasmer}, a popular
WebAssembly runtime, \emph{rust-url}, a URL parser, and \emph{rustls}, a TLS library.

Because these crates are primarily libraries, we evaluate a concrete binary target for each crate.
Specifically, we select one example or test target provided by the crate and use it for compilation
and execution. We build each target using Cargo's \texttt{debug} and \texttt{release} profiles and
repeat the build for each instrumentation configuration. \autoref{tab:eval_configs} shows nine
configurations that we use for our measurement.

For each build, we measure compilation time and final binary size. We focus on
code generation, where the inserted instrumentation is lowered and compiled, because this stage
accounts for most of the compilation overhead. \autoref{tab:eval_compilation} reports these results.
Due to the space limit, we only show the full results for the release profile (where the costs are
generally higher), and report only the averages for the debug profile. Under the MB configuration
(as described in \autoref{tab:eval_configs}), the release builds slow down by up to 72.4x with the
average size increase of 4.2x. Debug builds, on the other hand, slow down by up to 2.9X with the
average size increase of 5.2.

This variation is expected because \toolName instruments MIR, but much of the compilation work
happens after MIR is lowered. Additional metadata, such as concrete type ID and raw addresses for
places, increases this downstream cost. The effect also grows with the breadth of instrumentation,
as shown by the difference between local configurations (CF\textsuperscript{L} and
MB\textsuperscript{L}) and their complete counterpart configurations (CF and MB).

Overall, the compilation cost depends strongly on the amount of information collected and the amount
of code instrumented. For small and medium crates, the cost is generally manageable. For large
crates, comprehensive instrumentation can be expensive, making selective configuration important for
controlling compilation time and binary size.

\begin{table*}
\centering
\caption{Mean wall-clock time (sec) reported by benchmark programs when instrumented by \toolName using the configurations and when running Miri.}
\scriptsize
\label{tab:tab_eval_runtime}
\input{res_tbls_tab_eval_runtime}
\vspace{-1.5em}
\end{table*}

\mysubsubsection{Runtime Cost:}
We measure the runtime overhead of \toolName's instrumentation separately from the cost of any
specific analysis. This isolates the cost of executing inserted probes and constructing and
delivering DMIR events to the analysis. We expect this cost
to be substantial---\toolName does not merely record low-level events, but delivers structured
MIR-level information such as places, projections, concrete types, function identifiers, memory
addresses, and control flow decisions.

We use programs from \texttt{rust-perf}, the runtime performance benchmark suite used by the Rust
compiler project~\cite{rustc_perf}. We instrument all functions except those in the benchmarking library, so
that the measured overhead reflects the benchmarked program rather than the benchmarking
infrastructure. We then run each instrumented program with a no-op analysis, which receives events
but performs no analysis. The resulting slowdown therefore reflects the pure runtime cost of
instrumentation and DMIR delivery. \autoref{tab:tab_eval_runtime} reports the results.

The results show three main trends. First, overhead varies substantially across programs. Under the
MB configuration, slowdown ranges from 19x to 396x. Second, the cost depends strongly on the kind of
information delivered. Adding concrete type information increases overhead by 3.5x on average, and
adding raw place addresses increases overhead by 2.8x on average. This is expected because places
are frequent, type and address information can be large or costly to construct, and the inserted
code can inhibit later code-generation optimizations. Third, full MIR-level instrumentation is
expensive. Delivering rich DMIR events at runtime imposes a substantial cost even when the analysis
does no work.

To contextualize these costs, we discuss two forms of evidence from other dynamic tools. First, we
run the same programs with Miri. Miri serves a different purpose and is not a direct baseline for
\toolName, but it illustrates the cost of operating over MIR-level program information at runtime.
In our measurements, its overhead is 33x--163x higher than the MB configuration. Second, high
overhead is common even in instrumentation systems that target lower-level representations. Chen et
al. report pure instrumentation overhead of 13.9x, 128.3x, and 381.0x for SymSan, SymCC~\cite{poeplau_symbolic_2020}, and
SymQEMU~\cite{poeplau_symqemu_2021}, which target LLVM IR and binary representations for concolic
execution~\cite{chen_symsan_2022}. In this context, \toolName's runtime overhead is significant but
consistent with the cost of dynamically exposing rich execution information.

These results also show why configurability is necessary. The overhead varies substantially across
programs and configurations, and some information, such as concrete types and raw place addresses,
is especially expensive to deliver. A fixed instrumentation strategy would therefore be too rigid.
Some analyses need rich DMIR events over broad parts of the program, while others only need specific
events or specific code regions. \toolName addresses this by letting users configure both which
functions are instrumented and which entities are delivered. This allows analyses to trade off the
richness of the event stream against compilation time, runtime overhead, and binary size. This
design is consistent with prior work that prunes instrumentation based on analysis needs or static
properties of the target program~\cite{poeplau_symbolic_2020, min_erasan_2024}.

\subsection{RQ3: Faithfulness}

For RQ3, we evaluate whether \toolName's instrumentation preserves the behavior of instrumented
programs. Although the instrumentation pass does not modify the original MIR statements and checks
the consistency of the instrumented control flow, behavioral differences may still arise after
instrumentation. The inserted probes are compiled together with the program, MIR undergoes further
transformations during code generation, and \toolName changes parts of the linking process. Thus, we
empirically check whether instrumented programs still pass their original tests.

We use the same target projects, build profiles, and instrumentation configurations as in RQ2. For
each project, we run its test suite using \texttt{cargo test} in \texttt{release} profile with the
crate's default features, except for \texttt{wasmer}, where we use 17 of its example programs that
test its core API. We run the tests with a no-op analysis. This isolates the effect of instrumentation itself.

Across all configurations, the instrumented test suites pass except for 2 failures. In total, the
test suites contain 29,652 test cases, with some crates contributing thousands of tests and others
fewer than one hundred. The observed failures are due to unsupported cases in \toolName rather than
behavioral divergence in ordinary instrumented execution. In particular, \texttt{hashbrown} and
\texttt{rust-url} include one test case each that customizes panic handling, which the \toolName
implementation does not currently support. Apart from these known unsupported cases, the original
tests pass, providing empirical evidence that \toolName's instrumentation preserves program
behavior.

%% file: res_tbls_tab_eval_configs.tex
\begin{tabular}{lp{0.8\columnwidth}}
        \toprule
         \textbf{Name} & \textbf{Instrumentation Rules}\\
         \midrule
         OG& Excludes eveything. (no instrumentation) \\\hline
         CF&  Includes only control flow events, i.e., conditional branches and function calls.\\\hline
         MB&Includes all MIR events and their information except the lifetime events.\\\hline
         +L&Same as MIR, with the inclusion of lifetime events.\\\hline
                 +T& Same as MIR, with the inclusion of types of places.\\\hline
         +A&Same as MIR, with the inclusion of raw addresses for places.\\\hline
         +FA&Same as MIR, with the inclusion of raw addresses of functions.\\\hline
         ALL&Includes every entity supported, i.e., MIR with inclusion all mentioned above.\\\hline
         (C)\textsuperscript{L}&Same as the configuration C, but only including local functions, i.e., excluding the functions in the dependencies.\\ \bottomrule
\end{tabular}

%% file: res_tbls_tab_eval_compilation.tex
\begin{tabular}{|ll|r||r|r|r|r|r|r|r|r|r|r|}
\hline
 &  & \textbf{OG} & \textbf{CF} & \textbf{CF\textsuperscript{L}} & \textbf{MB} & \textbf{MB\textsuperscript{L}} & \textbf{+L} & \textbf{+T} & \textbf{+A} & \textbf{+FA} & \textbf{ALL} \\
 \textbf{Crate} & \textbf{Metric} &  &  &  &  &  &  &  &  &  &  \\
 \hline
 
 \hline
 
 \multirow[m]{2}{*}{\code{bitflags}} & CG (Rate) & 1.0s & 1.0s (1.1x) & 1.2s (1.2x) & 2.3s (2.4x) & 1.2s (1.3x) & 2.8s (2.9x) & 4.3s (4.4x) & 4.8s (5.0x) & 4.2s (4.3x) & 15s (15.4x) \\
  & Size (Rate) & 1MB & 1MB (1.3x) & 1MB (1.0x) & 1MB (2.7x) & 1MB (1.0x) & 2MB (3.3x) & 2MB (4.7x) & 3MB (5.4x) & 2MB (4.4x) & 7MB (14.3x) \\
\cline{1-12}
 \multirow[m]{2}{*}{\code{flate2}} & CG (Rate) & 2.0s & 2.2s (1.1x) & 2.0s (1.0x) & 5.5s (2.7x) & 2.1s (1.0x) & 6.5s (3.2x) & 9.0s (4.5x) & 10s (5.0x) & 8.6s (4.2x) & 35s (17.2x) \\
  & Size (Rate) & 1MB & 1MB (1.6x) & 1MB (1.1x) & 3MB (4.1x) & 1MB (1.2x) & 4MB (5.3x) & 5MB (7.1x) & 5MB (8.0x) & 4MB (6.3x) & 15MB (22.9x) \\
\cline{1-12}
 \multirow[m]{2}{*}{\code{crossterm}} & CG (Rate) & 2.9s & 4.2s (1.5x) & 2.8s (1.0x) & 8.8s (3.1x) & 3.3s (1.1x) & 11s (4.0x) & 16s (5.5x) & 15s (5.3x) & 12s (4.1x) & 2m (38.6x) \\
  & Size (Rate) & 1MB & 1MB (1.7x) & 1MB (1.1x) & 4MB (4.6x) & 1MB (1.4x) & 6MB (6.7x) & 7MB (8.7x) & 7MB (8.8x) & 5MB (6.5x) & 28MB (33.0x) \\
\cline{1-12}
 \multirow[m]{2}{*}{\code{rustls}} & CG (Rate) & 9.8s & 11s (1.1x) & 10s (1.1x) & 27s (2.7x) & 16s (1.6x) & 32s (3.2x) & 48s (4.9x) & 47s (4.8x) & 36s (3.7x) & 2m (12.0x) \\
  & Size (Rate) & 12MB & 13MB (1.1x) & 12MB (1.1x) & 21MB (1.8x) & 15MB (1.3x) & 24MB (2.1x) & 32MB (2.8x) & 32MB (2.8x) & 25MB (2.2x) & 65MB (5.6x) \\
\cline{1-12}
 \multirow[m]{2}{*}{\code{hashbrown}} & CG (Rate) & 12s & 13s (1.1x) & 11s (1.0x) & 33s (2.8x) & 12s (1.0x) & 40s (3.4x) & 60s (5.2x) & 62s (5.4x) & 43s (3.7x) & 3m (17.6x) \\
  & Size (Rate) & 2MB & 4MB (1.8x) & 3MB (1.1x) & 13MB (5.5x) & 3MB (1.2x) & 17MB (7.1x) & 26MB (10.8x) & 26MB (10.7x) & 18MB (7.4x) & 74MB (30.4x) \\
\cline{1-12}
 \multirow[m]{2}{*}{\code{rust-url}} & CG (Rate) & 13s & 16s (1.2x) & 14s (1.0x) & 39s (2.9x) & 16s (1.2x) & 51s (3.8x) & 72s (5.4x) & 72s (5.4x) & 52s (3.9x) & 10m (44.2x) \\
  & Size (Rate) & 3MB & 6MB (1.8x) & 4MB (1.1x) & 17MB (5.1x) & 5MB (1.4x) & 24MB (7.1x) & 34MB (10.0x) & 33MB (9.8x) & 23MB (6.8x) & 110MB (32.9x) \\
\cline{1-12}
 \multirow[m]{2}{*}{\code{bat}} & CG (Rate) & 30s & 37s (1.3x) & 30s (1.0x) & 95s (3.2x) & 31s (1.1x) & 2m (4.0x) & 3m (6.3x) & 3m (6.0x) & 2m (4.7x) & 15m (30.4x) \\
  & Size (Rate) & 5MB & 9MB (1.6x) & 5MB (1.0x) & 23MB (4.3x) & 6MB (1.1x) & 29MB (5.4x) & 47MB (8.8x) & 42MB (7.7x) & 31MB (5.8x) & 135MB (25.1x) \\
\cline{1-12}
 \multirow[m]{2}{*}{\code{wasmer}} & CG (Rate) & 2m & 2.0h (60.4x) & 2m (1.0x) & 2.4h (72.4x) & 2m (1.1x) & 2.2h (66.3x) & 2.3h (71.4x) & 2.4h (74.6x) & 2.6h (80.1x) & 3.3h (102.6x) \\
  & Size (Rate) & 24MB & 43MB (1.8x) & 25MB (1.0x) & 125MB (5.3x) & 28MB (1.2x) & 155MB (6.6x) & 248MB (10.5x) & 241MB (10.2x) & 172MB (7.3x) & 613MB (26.0x) \\
\cline{1-12}
 \multirow[m]{2}{*}{\textbf{Average}} & CG Rate & 1.0 & 8.6 (±19.6) & 1.0 (±0.1) & 11.5 (±23.0) & 1.2 (±0.2) & 11.4 (±20.8) & 13.4 (±21.9) & 13.9 (±22.9) & 13.6 (±25.1) & 34.8 (±27.9) \\
  & Size Rate & 1.0 & 1.6 (±0.2) & 1.1 (±0.0) & 4.2 (±1.2) & 1.2 (±0.1) & 5.4 (±1.8) & 7.9 (±2.7) & 7.9 (±2.5) & 5.9 (±1.6) & 23.8 (±8.9) \\
\cline{1-12}
 \textbf{Average} & CG Rate & 1.0 & 1.1 (±0.0) & 1.0 (±0.0) & 2.1 (±0.5) & 1.1 (±0.1) & 2.6 (±0.6) & 6.0 (±2.0) & 3.2 (±0.8) & 3.2 (±0.5) & 17.4 (±6.7) \\
 \textbf{(Debug)} & Size Rate & 1.0 & 2.4 (±0.4) & 1.1 (±0.1) & 5.2 (±1.0) & 1.4 (±0.3) & 6.3 (±1.4) & 13.8 (±3.3) & 8.2 (±1.7) & 7.0 (±1.3) & 34.9 (±10.5) \\
\cline{1-12}
\end{tabular}

%% file: res_tbls_tab_eval_runtime.tex
\begin{tabular}{|l|r||r|r|r|r|r|r|r!{\vrule width 0.75pt}r|}
\hline
 & \textbf{OG} & \textbf{CF} & \textbf{MB} & \textbf{+L} & \textbf{+T} & \textbf{+A} & \textbf{+FA} & \textbf{ALL} & \textbf{Miri} \\
\textbf{Benchmark} &  &  &  &  &  &  &  &  &  \\
\hline

\hline
\code{brotli-compress} & 0.42 & 5.65\text{ }(13.4x) & 33.9\text{ }(80.5x) & 71.7\text{ }(170x) & 146\text{ }(345x) & 101\text{ }(240x) & 32.8\text{ }(77.9x) & 515\text{ }(1,223x) & 17,129 \\
\code{brotli-decompress} & 0.20 & 2.63\text{ }(13.3x) & 14.8\text{ }(74.6x) & 21.0\text{ }(106x) & 42.2\text{ }(213x) & 34.3\text{ }(173x) & 14.4\text{ }(72.7x) & 84.1\text{ }(425x) & 4,109 \\
\code{bufreader_snappy} & 0.06 & 0.30\text{ }(5.20x) & 2.19\text{ }(38.5x) & 3.64\text{ }(64.0x) & 5.57\text{ }(97.7x) & 4.97\text{ }(87.2x) & 2.14\text{ }(37.5x) & 18.3\text{ }(321x) & 2,994 \\
\code{css-parse-fb} & 0.12 & 2.57\text{ }(21.8x) & 13.1\text{ }(111x) & 30.7\text{ }(261x) & 38.6\text{ }(328x) & 33.1\text{ }(281x) & 12.9\text{ }(109x) & 129\text{ }(1,096x) & 10,466 \\
\code{fmt-debug-derive} & 0.01 & 0.11\text{ }(12.4x) & 0.43\text{ }(49.1x) & 1.38\text{ }(159x) & 2.40\text{ }(277x) & 1.97\text{ }(226x) & 0.43\text{ }(49.1x) & 10.1\text{ }(1,162x) & 743 \\
\code{fmt-write-str} & 0.02 & 0.22\text{ }(10.8x) & 0.90\text{ }(44.1x) & 2.63\text{ }(129x) & 7.87\text{ }(387x) & 4.96\text{ }(244x) & 0.89\text{ }(43.8x) & 31.7\text{ }(1,558x) & 1,390 \\
\code{hashmap_find} & 0.02 & 0.43\text{ }(18.4x) & 0.96\text{ }(41.3x) & 1.15\text{ }(49.3x) & 1.68\text{ }(72.2x) & 1.73\text{ }(74.4x) & 0.74\text{ }(31.8x) & 16.2\text{ }(697x) & 704 \\
\code{hashmap_find_misses} & 0.00 & 0.17\text{ }(49.2x) & 0.49\text{ }(140x) & 0.70\text{ }(203x) & 1.01\text{ }(291x) & 1.01\text{ }(291x) & 0.43\text{ }(124x) & 5.19\text{ }(1,490x) & 457 \\
\code{hashmap_insert} & 0.03 & 0.18\text{ }(6.25x) & 0.53\text{ }(19.0x) & 0.72\text{ }(25.8x) & 1.03\text{ }(36.7x) & 1.01\text{ }(35.9x) & 0.43\text{ }(15.2x) & 12.7\text{ }(454x) & 443 \\
\code{hashmap_iterate} & 0.00 & 0.04\text{ }(17.4x) & 0.25\text{ }(106x) & 0.36\text{ }(155x) & 0.57\text{ }(244x) & 0.53\text{ }(228x) & 0.24\text{ }(101x) & 1.28\text{ }(546x) & 174 \\
\code{hashmap_remove} & 0.03 & 0.48\text{ }(15.5x) & 1.21\text{ }(38.7x) & 1.64\text{ }(52.5x) & 2.42\text{ }(77.7x) & 2.31\text{ }(74.0x) & 1.08\text{ }(34.5x) & 27.5\text{ }(882x) & 1,042 \\
\code{nom-json} & 0.09 & 5.37\text{ }(60.6x) & 28.6\text{ }(323x) & 64.5\text{ }(728x) & 127\text{ }(1,432x) & 90.6\text{ }(1,022x) & 28.3\text{ }(319x) & 505\text{ }(5,697x) & 43,534 \\
\code{pinky-nes15} & 0.17 & 16.8\text{ }(98.1x) & 67.9\text{ }(396x) & 125\text{ }(731x) & 198\text{ }(1,152x) & 168\text{ }(980x) & 67.7\text{ }(395x) & 529\text{ }(3,087x) & 16,047 \\
\code{regex-capture-1} & 0.01 & 0.25\text{ }(30.3x) & 1.74\text{ }(211x) & 3.91\text{ }(472x) & 8.83\text{ }(1,066x) & 6.54\text{ }(790x) & 1.74\text{ }(210x) & 32.3\text{ }(3,897x) & 2,457 \\
\code{regex-search-1} & 0.00 & 0.05\text{ }(15.1x) & 0.92\text{ }(268x) & 1.88\text{ }(548x) & 3.42\text{ }(997x) & 2.63\text{ }(767x) & 0.90\text{ }(263x) & 9.27\text{ }(2,703x) & 699 \\
\hline
\end{tabular}

%% file: parts_s7-related.tex
\section{Related Work}

\mysubsubsection{Dynamic Analysis Frameworks:}
Dynamic analysis frameworks span languages and representations.
DynInst~\cite{williams_dyninst_2016},
DynamoRIO~\cite{bruening_infrastructure_2003}, Pin~\cite{luk_pin_2005},
Valgrind~\cite{nethercote_valgrind_2007}, and Frida~\cite{frida} operate at the
binary level.
RoadRunner~\cite{flanagan_roadrunner_2010} provides a Java event-stream API.
DiSL~\cite{marek_disl_2012} defines Java analyses through a domain-specific
language. Wasabi~\cite{lehmann_wasabi_2019} and
DynaPyt~\cite{eghbali_dynapyt_2022} make selective instrumentation central for
WebAssembly and Python. These systems inspired \toolName's interface design and
configurable instrumentation. Our challenges differ in preserving MIR semantics
and exposing them through DMIR callbacks.

\mysubsubsection{Dynamic Analysis for Rust:}
Miri~\cite{jung_miri_2026} and RuDyna~\cite{deng_rudyna_2025} are closest to \toolName. 
Miri interprets programs for checking correctness and undefined behavior.
RuDyna focuses on inserting hooks for MIR events, demonstrated with small use cases. In contrast, \toolName provides DMIR as a
higher-level interface and implements the instrumentation techniques needed to deliver MIR
and runtime information as an event stream, with first-class support for constructs such as
places and operands. \toolName also addresses the engineering needs for high-coverage
instrumentation, demonstrated by our three significant use cases.
Other Rust dynamic analysis tools work at the level of LLVM-IR or
native binaries, such as porting  KLEE~\cite{cadar_klee_2008} to Rust (RVT~\cite{rvt, zhang_broadly_2024}), porting SymCC~\cite{poeplau_symbolic_2020}
to Rust (SymRustc)~\cite{tuong_symrustc_2023}), and enabling LLVM sanitizers~\cite{rustc_sanitizers}.

Although applicable to Rust, they lose or must reconstruct Rust-level semantic
information, as done for instance in ERASan~\cite{min_erasan_2024}.

\mysubsubsection{Rust Analysis, Verification, and Bugs:}
Kani~\cite{kani} performs model checking,
Aeneas~\cite{ho_aeneas_2022} and Charon~\cite{ho_charon_2025} support formal
verification, and RAPx~\cite{rapx} provides a suite of static analyses.
Flowistry~\cite{crichton_modular_2022} and Cocoon~\cite{lamba_cocoon_2024}
target information flow, SafeDrop~\cite{cui_safedrop_2023} and
Rudra~\cite{bae_rudra_2021} target memory-safety issues, and
Rupta~\cite{li_context-sensitive_2024} revisits pointer analysis for Rust.
Empirical studies of Rust bugs and vulnerabilities~\cite{zhang_beyond_2024, robatishirzad_study_2024,
zheng_closer_2024, mccormack_study_2025} motivate the need for runtime facts,
e.g., ownership and places.

%% file: parts_s9-conclusion.tex
\section{Conclusion}

In this paper, we presented \toolName, a Rust-native framework for building dynamic analyses.
\toolName targets MIR to capture Rust's semantics and exposes them through DMIR, an event-driven
interface that augments MIR-level operations with runtime information. Internally, \toolName
combines MIR instrumentation, runtime support, and configurability to bridge compile-time MIR and
runtime execution. Our evaluation shows that \toolName supports substantial analyses, faithfully
delivers Rust semantics and execution events, and provides configurability to manage the measurable
costs of MIR-level instrumentation. Overall, \toolName demonstrates that MIR is a practical
foundation for Rust-specific dynamic analysis.

\mysubsubsection{Acknowledgment:} This paper has used Claude (Sonnet 4.6) and
Copilot (GPT 5.5) for grammar, smoothness, and clarity.

%% file: parts_s_-references.tex
\bibliographystyle{IEEEtran}
\bibliography{main}